\documentclass[9pt,showpacs]{revtex4-1}

\usepackage{epsfig}
\input epsf.tex
\usepackage{amssymb}
\usepackage{amsmath}
\usepackage{amsthm}
\usepackage{amsopn}

\def\comment#1{}

\def\beq{\begin{equation}}
\def\eeq{\end{equation}}
\def\bea{\begin{eqnarray}}
\def\eea{\end{eqnarray}}

\usepackage{ulem}
\usepackage{cancel}
\usepackage{color}

\begin{document}

\title{ Generation of CMB B-mode Polarization from Circular Polarization}

\author{Rohoollah Mohammadi$^{1,2}$}
\email[]{rmohammadi-AT-ipm.ir}

\author{Moslem Zarei$^{3,4}$}
\email[]{m.zarei-AT-cc.iut.ac.ir}

\affiliation{$^1$Iran Science and Technology Museum (IRSTM), PO BOX: 11369-14611, Tehran, Iran.}
\affiliation{$^2$School of physics, Institute for research in fundamental sciences (IPM), Tehran, Iran.}

\affiliation{$^3$ Department of Physics, Isfahan University of
Technology, Isfahan 84156-83111, Iran}

\affiliation{$^4$ School of Astronomy, Institute for Research in Fundamental
Sciences (IPM), P. O. Box 19395-5531, Tehran, Iran}

\date{\today}

\begin{abstract}

In this work we consider non-zero circular polarization for the CMB radiation as a result of new interactions. We then rewrite the Boltzmann equations for the Stokes parameters $Q$, $U$ and $V$ and show that the circular polarization can generate the B-mode polarization even if no tensor perturbations are present.

\end{abstract}

%\pacs{13.15.+g,34.50.Rk,13.88.+e}

\maketitle
%\newpage

\section{Introduction}

The precise measurement of cosmic microwave background (CMB) polarization will play a major role in discriminating between inflationary models. The CMB polarization patterns can be decomposed into what are called E-modes and B-modes \cite{Zaldarriaga:1996xe,Kamionkowski:1996ks,Kosowsky:1994cy,Hu:1997hp}. It is well known that the B-mode polarization is generated by tensor perturbations \cite{Zaldarriaga:1996xe,Kamionkowski:1996ks,Kosowsky:1994cy,Hu:1997hp}. Inflationary models predict the scalar power spectrum, $\mathcal{P}_{s} (k)$, and the tensor power spectrum, $\mathcal{P}_{t} (k)$, which are naturally generated from quantum fluctuations of inflaton field during inflation \cite{Guth:1980zm,Linde:1981mu,Linde:1983gd,Bardeen:1983qw,Kosowsky:1995aa}. The amplitude of tensor perturbation is characterized by the tensor-to-scalar ratio, $r=\mathcal{P}_{t}/\mathcal{P}_{s}$. This ratio usually are given by comparing E-mode and B-mode polarization of CMB. A joint analysis of the BICEP2/Keck Array and Planck favors solutions without gravity waves and have reported the upper limit on the tensor-to-scalar ratio is $r < 0.11$ \cite{Planck}.

\begin{figure}
  \includegraphics[width=4in]{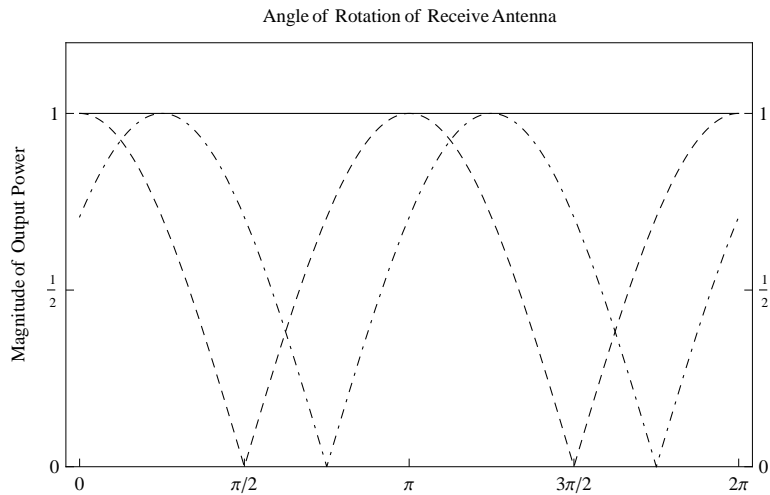}\\
  \caption{The magnitude of output power  as function the angle of rotation of a receive antenna is shown in case of
  when a receive antenna detects a linearly polarized  source beam (dashed and dot-dashed line) or  circular polarized  source beam (solid line).
  }\label{fig1}
\end{figure}

The tensor modes supply the non-vanishing values for the off-diagonal components of CMB polarization matrix. The B-mode is generated when the off-diagonal elements of polarization matrix are non-zero. Accordingly, the future detection B-mode polarization signals acts as fingerprint of gravitational waves. We assert that instead of tensor perturbations there are other mechanism generating the B-mode polarization \cite{Bonvin:2014xia,Khodagholizadeh:2014nfa,Moss:2014cra,Lizarraga:2014eaa}. In this work we also show that the circular polarization also can act as a source for generating B-modes. The circular polarization in polarization matrix is parameterized by Stokes parameter V. Usually it is assumed that at the last scattering surface the Thomson (Compton) scattering does not generate the intrinsic circular polarization of CMB and hence one can input $V=0$ in the polarization matrix. However, in recent years it have been shown that the primordial circular polarization can be generated due to some effects for example photon-neutrino interaction or the presence of background
magnetic fields \cite{Mohammadi:2013,Alexander:2008fp,Bavarsad:2009hm,Giovannini,Motie:2011az,De:2014qza,Sawyer:2014maa}. Though, there is no plan to detect the V-mode of CMB in future, the MIPOL experiment has reported an upper bound ranging between $5.0\times 10^{-4}$ and $0.7\times 10^{-4}$ at angular scales between $8^{\circ}$ and $24^{\circ}$ on the degree of CMB circular polarization \cite{Mainini:2013mja}.\\
 \indent
In this sense, one can consider the non-zero circular polarization parameter, V, in the polarization matrix. We can study the effects of circular polarization on the B-mode polarization in two aspects. \\
 \indent
First, suppose that a source beam is vertically polarized, and that a receive antenna is also vertically linearly polarized such that the direction of linear polarization of the source beam and the antenna are matched. We set output power is set equal to one for simplicity, then the normalized output of our experiment, as a function of the rotation angle of the receive antenna, would look something like the dashed line shown in Fig.(\ref{fig1}).
 In this figure the dot-dashed line shows the case that the direction of receive antenna is different from linearly polarized source beam by $\pi/4$.
Suppose now that the source beam was radiating a right or left hand circularly polarized wave. If the source beam was subjected to the same measurement as above, the normalized output power would resemble that of Fig.(\ref{fig1}) as solid line.
This means that a receive antenna with a fixed direction records the total power of circular polarization (independent of direction) as it is recording some contribution from linear polarization dependent of its direction with respect to our receive antenna. In other words, independence of the direction of the receive antenna, the receive antenna can not cancel the measurement of circular polarization in contrast to linear polarization. In BICEP2 experiment, there are
partial view of one BICEP2 dual-polarization pixel is shown in \cite{BICEP2}. The vertically oriented slots are sensitive to horizontal polarization and form the antenna network for the A detector, while the horizontally oriented slots receive vertical polarization and are fed into the B detector. In this way, the A and B detectors have orthogonal polarizations but are spatially co-located and form beams that are coincident on the sky. This view corresponds to a bore sight angle of $90^{\circ}$. At bore sight angle of $0^{\circ}$, the A detectors receive vertical polarization and the B detectors receive horizontal polarization \cite{BICEP2}. As discussed in previous paragraph, the both vertically and horizontally antennas of BICEP2 experiment measure not only linear polarization of CMB but also the circular polarization one. As a result, the E- and B- mode polarization contain the linear and circular polarizations both which can affect on result of the value of r-parameter. This means at first the contribution of circular polarization on B- mode should be distinguished from its contribution from linear polarization due to tensor mode and then by using the correct value of B- mode due to
tensor mode, one can estimate the value of r-parameter.\\
 \indent
Second, the circular polarization is contributed as a source term for the Stokes parameter U upon considering non-zero value for V. Usually, the right hand side of the Boltzmann equations, the collision term, is computed by assuming the conventional Thomson scattering. The other interactions are not contributes because of the smallness of their scattering cross sections in compare with the Thomson cross-section. If we take into account the Thomson scattering only, as it is well known and we will see in the following, the B-mode polarization cannot be generated in case of scalar perturbations. We will discuss that this property can change by employing the interactions generating the circular polarization. We generally show that the non-zero circular polarization mode generating through new interactions can be a sufficient source to generate B-mode polarization.

The paper is organized as follows:
In section II We review the derivation of the standard coupled Boltzmann equations of the CMB photons for the scalar perturbations. In section III we consider the new interactions between CMB photons and electrons generating the circular polarization and discuss their effects on the Boltzmann equations. In this section we derive the new form of Boltzmann equations and show that a non-zero V parameter can lead to a non-zero B-mode.

\section{ Boltzmann Equations of the CMB Radiation}

The CMB Polarization is described in terms of Stokes parameters $U $ ,$ Q $ ,$ V $. They are related to the matrix polarization $\mathbf{P}$ as follows ( see e.g. \cite{Hu:1997hp} and references there in)
\bea
\mathbf{P}=\Theta \sigma_0+Q\sigma_3+ U \sigma_1+V \sigma_2~,
\eea
where $\sigma$'s are the Pauli matrices, $\Theta $ is temperature perturbation, $Q$ and $U$ describe the linear polarization and
the circular polarization corresponds to $ Q=U=0 $, $ V\neq 0 $. It is convenient to rewrite the polarization matrix in terms of $\sigma_{\pm}=(\sigma_3\mp \sigma_1)/2 $ in the following form
 \bea
\mathbf{P}=\Theta \sigma_0+(Q+iU)\sigma_+\,+ (Q-iU) \sigma_-+V \sigma_2~,
\eea
Accordingly, the $Q\pm iU$ now transform like spin 2 objects under the rotation. We define the 4-vector
\beq
P\equiv P^{\mu}=(\Theta, Q+iU, Q-iU,V)~,
\eeq
so that the evolution of $P$ is characterized using the Boltzmann equation
\beq
\frac{\partial}{\partial\eta}P+n^{i}\nabla_{i}P=C[P]+G[h_{\mu\nu}]~, \label{Boltz0}
\eeq
where $\eta$ is the conformal time, $G[h_{\mu\nu}]$ is denoting the gravitational red-shift due to metric perturbations and $C[P]$ is the collision term calculated as follows. It is known that the Thomson scattering transforms the polarization vector as \cite{Hu:1997hp,Chandrasekhar}
\beq
\left(
  \begin{array}{c}
    \Theta\\
    Q+iU \\
    Q-iU \\
    V \\
  \end{array}
\right)^{\textrm{scatt.}}
=\frac{3}{4}\left(
       \begin{array}{cccc}
         \cos^{2}\beta+1 & -\frac{1}{2}\sin^{2}\beta & -\frac{1}{2}\sin^{2}\beta & 0 \\
         -\frac{1}{2}\sin^{2}\beta & \frac{1}{2}(\cos^{2}\beta+1)^{2} & \frac{1}{2}(\cos^{2}\beta-1)^{2} & 0 \\
          -\frac{1}{2}\sin^{2}\beta &\frac{1}{2}(\cos^{2}\beta-1)^{2} & \frac{1}{2}(\cos^{2}\beta+1)^{2} & 0 \\
         0 & 0 & 0 & \cos\beta \\
       \end{array}
     \right)\left(
  \begin{array}{c}
    \Theta \\
    Q+iU \\
    Q-iU \\
    V \\
  \end{array}
\right)~,\label{scattering matrix0}
\eeq
where $\beta$ is the scattering angle. We now perform a coordinate rotations in order to transform to a frame in which the wave vector $\mathbf{k}$ oriented along the $\hat{\mathbf{z}}$ direction. Such transformations change the scattering matrix into the form
\beq
S_T(\mathbf{n},\widetilde{\mathbf{n}})=\frac{4\pi}{10}\sum_{m=-2}^{2}\left(
                        \begin{array}{cccc}
                          Y_{2}^{\widetilde{m}} Y_{2}^{m} & -\sqrt{\frac{3}{2}} \,_{2}Y_{2}^{\widetilde{m}} Y_{2}^{m} & -\sqrt{\frac{3}{2}} \,_{-2}Y_{2}^{\widetilde{m}} Y_{2}^{m} & 0 \\
                          -\sqrt{6}Y_{2}^{\widetilde{m}} \,_{2}Y_{2}^{m} & 3\,_{2}Y_{2}^{\widetilde{m}} \,_{2}Y_{2}^{m}  & 3 \,_{-2}Y_{2}^{\widetilde{m}} \,_{2}Y_{2}^{m}  & 0 \\
                          -\sqrt{6}Y_{2}^{\widetilde{m}}  \,_{2}Y_{2}^{m}  & 3\,_{2}Y_{2}^{\widetilde{m}} \,_{-2}Y_{2}^{m}  & 3 \,_{-2}Y_{2}^{\widetilde{m}} \,_{-2}Y_{2}^{m}  & 0 \\
                          0 & 0& 0 &\frac{5}{2} Y_{1}^{\widetilde{m}} Y_{1}^{m}  \\
                        \end{array}\label{scattering matrix}
                      \right)~,
\eeq
It is therefore straightforward to rewrite the collision term $C[T]$ in the electron rest frame in the following form \cite{Hu:1997hp}
\bea
C[P]=\tau' I(\mathbf{n})+\tau'\int \frac{d\widetilde{\mathbf{n}}}{4\pi}S_T(\mathbf{n},\widetilde{\mathbf{n}})P(\widetilde{\mathbf{n}})~, \label{scattering}
\eea
where $\tau'$ is the optical depth and the vector $I$ describes the isotropization of distribution in the electron rest frame given by
\beq
I(\mathbf{n})=P(\mathbf{n})-\left(\int \frac{d\widetilde{\mathbf{n}}}{4\pi}\Theta(\widetilde{\mathbf{n}})+\hat{\mathbf{n}}\cdot \mathbf{v}_{B},0,0,0\right)~,
\eeq
where  the term $\hat{\mathbf{n}}\cdot \mathbf{v}_{B}$ contributes due to the Doppler boosting from the electron rest frame into the background frame. As well as, the cosmological redshift due to stretching of the the spacial metric, frame dragging and time dilation are the three effects of gravitational perturbations on the CMB photon trajectory. It is possible to summarize these effects into the the gravitational interaction vector $G[h_{\mu\nu}]$ which for scalars is given by \cite{Hu:1997hp}
\beq
G[h_{\mu\nu}]=\left(\frac{1}{2}n^{i}n^{j}\dot{h}_{ij}+\frac{1}{2}n^{i}\nabla _{i}h_{00},0,0,0\right)~, \label{gravityvector}
\eeq
where here dot denotes derivative with respect to time. In Newtonian gauge one can choose
\beq
h_{00}=2\Phi e^{i\mathbf{k}\cdot \mathbf{x}}~, \:\:\:\:\:\textrm{and}\:\:\:\:\: h_{ij}=2\Psi e^{i\mathbf{k}\cdot \mathbf{x}}\delta_{ij}~. \label{h}
\eeq
As it is clear from \eqref{gravityvector} the scalar perturbations of the metric cannot alter the polarization parameters. However, the tensor modes can change the polarization parameters $U$ and $Q$ although, the circular polarization parameter $V$ remains unchanged. While we ignore the tensor perturbation, we use the new variables $\Delta_{T}^{S}$, $\Delta_{Q}^{S}$, $\Delta_{U}^{S}$ and $\Delta_{V}^{S}$ \cite{Zaldarriaga:1996xe,Kamionkowski:1996ks,Kosowsky:1994cy,Hu:1997hp} to assign the temperature and polarization anisotropies respectively with the superscript S denoting the scalar modes. Now by using \eqref{Boltz0} and \eqref{scattering}, inserting \eqref{h} into \eqref{gravityvector} and after some straightforward algebra one can arrive in the following well-known Boltzmann equations describing the CMB anisotropies \cite{Zaldarriaga:1996xe,Kamionkowski:1996ks,Kosowsky:1994cy,Hu:1997hp}
\beq
\Delta_{T}^{'S}+ik\mu\left(\Delta_{T}^{S}+\Psi\right)-\Phi'=\tau'\left(\Delta_{T0}^{S}-\Delta_{T}^{S}+ik\mu v_{B}+\frac{1}{2}P_{2}\,\Pi\right)~,
\eeq
\beq
\Delta_{Q}^{'S}+ik\mu\Delta_{Q}^{S}=\tau'\left(-\Delta_{Q}^{S}+\frac{1}{2}[1-P_{2}]\,\Pi\right)~,
\eeq
\beq
\Delta_{U}^{'S}+ik\mu\Delta_{U}^{S}=-\tau'\Delta_{U}^{S}~, \label{u1}
\eeq
\beq
\Delta_{V}^{'S}+ik\mu\Delta_{V}^{S}=-\tau'\left(\Delta_{V}^{S}+\frac{3}{2}i\mu \Delta_{V1}^{S}\right)~,  \label{v1}
\eeq
where $P_{l}(\mu)$ is the Legendre function, $\mu=\hat{\mathbf{n}}\cdot\hat{\mathbf{k}}$ and $\Pi=\Delta_{T2}^{S}+\Delta_{Q2}^{S}+\Delta_{Q0}^{S}$ is given after integrating over the solid angle $\widetilde{\mathbf{n}}$ in \eqref{scattering} and by using the properties of the spherical harmonic. Here and throughout the paper prime in the Boltzmann equations denotes a derivative with respect to the conformal time. These system of equations can be solved numerically in order to calculate the CMB angular power spectrum $C_{l}^{XY}$. At this stage, one important result is that the equations \eqref{u1} and \eqref{v1} are not coupled to each other and to the other equations. This means that the scalar perturbations cannot generate the $U$ and the $V$ polarization modes. Taking into account the tensor perturbations contributes a source term on the right hand side of \eqref{u1} which leads to non-zero values for the $U$ parameter and the B-mode polarization as a consequence. However, the tensor modes do not contribute new terms in equation \eqref{v1} and hence the main conclusion about the circular polarization remains unchanged.

\section{New interactions and the generation of B-modes}

As we described above the conventional electron-photon interaction or Thomson (Compton) scattering cannot generate the U and V polarization parameters. However, it is known that the new interactions may produce the V parameter \cite{Mohammadi:2013,Alexander:2008fp,Bavarsad:2009hm,Giovannini,Motie:2011az,De:2014qza,Sawyer:2014maa}. Let's assume $V\neq 0$ due to some mechanisms such as neutrino-photon scattering via loop corrections \cite{Khodagholizadeh:2014nfa} or primordial magnetic field \cite{Bavarsad:2009hm,Giovannini}. Here, we consider the new interactions generating circular polarization in a general manner. At the first order, the new interaction can generally change the evolution of Stokes parameters Q, U and V. In general, one can parameterize the effects of new interactions by extending the scattering matrix in equation \eqref{scattering matrix0} in the following form

\beq
S=\frac{3}{4}\left(
       \begin{array}{cccc}
         \cos^{2}\beta+1 & -\frac{1}{2}\sin^{2}\beta & -\frac{1}{2}\sin^{2}\beta & 0 \\
         -\frac{1}{2}\sin^{2}\beta & \frac{1}{2}(\cos^{2}\beta+1)^{2} & \frac{1}{2}(\cos^{2}\beta-1)^{2} & 0 \\
          -\frac{1}{2}\sin^{2}\beta &\frac{1}{2}(\cos^{2}\beta-1)^{2} & \frac{1}{2}(\cos^{2}\beta+1)^{2} & 0 \\
         0 & 0 & 0 & \cos\beta \\
       \end{array}
     \right)
     +\frac{3}{4}\left(
       \begin{array}{cccc}
         0 &0 & 0 & 0 \\
         0 & \delta_{22} & \delta_{23} & \delta_{24} \\
         0 &\delta_{32} & \delta_{33} & \delta_{34} \\
         0 & \delta_{42} & \delta_{43} &0 \\
       \end{array}
     \right)~,\label{correction matrix}
\eeq
where the $\delta_{ij}$ corrections are appeared due to the new interaction terms which change the collision terms  in the following form
\beq
C[P]=C_{T}\left[\epsilon_s\cdot\epsilon_{\tilde{s}}\right]+
C_{\textrm{New}}\left[\epsilon_s,\epsilon_{\tilde{s}},k,q\right]\,~,
\eeq
where the $C_{T}$ is the collision term due to the Thomson scattering and the $C_{\textrm{New}}$ term is corresponding to the possible new interactions. Here $\epsilon_{s}$ denotes the polarization vector of CMB photons and $q$ and $k$ are the four momentum of interacting particles. The first row and the first column of the correction matrix in \eqref{correction matrix} consists entirely of zeros. This consideration comes from the nature of new interactions and their leading terms. In following, by introducing the quantum Boltzmann equation for the time evaluation of CMB matrix density, $\mathbf{P}$, we discuss this point. In our case, the time evaluation of $\mathbf{P}$ due to the new interaction Hamiltonian, $H_I$, can be simply written as \cite{Kosowsky:1994cy}
\begin{eqnarray}
\frac{d}{dt}\mathbf{P} =[H_I,\mathbf{P}]+\int dt [H_I,[H_I,\mathbf{P}]],\label{B1}
\end{eqnarray}
where the leading non-trivial contribution of new interactions comes from the forward scattering amplitude  $[H_I,\mathbf{P}]$
whereas the second term, $\int dt [H_I,[H_I,\mathbf{P}]]$, corresponding to the higher order new interaction terms, gives the scattering cross section which is highly subdominant compared to the Thomson scattering. Hence, we neglect this term for the rest. Note that the forward scattering of CMB photons with any particle can not change their direction and consequently can not change the CMB anisotropy $\Delta_{T}$. In the other word, one can set $\delta_{0\mu}=\delta_{\mu0}=0$ in \eqref{correction matrix}. For the sake of clarity, here we mention some examples. In \cite{Bavarsad:2009hm}, the effects of Compton forward scattering due to the electrodynamic
sector of the standard model is extended by the Lorentz non-invariant operators as well as the
non-commutativity effects. As it has been shown in \cite{Bavarsad:2009hm}, the $\delta_{ij}$ elements contain Lorentz non-invariant and non-commutativity parameters and the non-vanishing components are given by
$\{\delta_{43},\delta_{42},\delta_{24},\delta_{34}\}\neq0$. In letter \cite{Khodagholizadeh:2014nfa}, by considering forward scattering between the CMB photons and the cosmic neutrinos background, the first order of quantum Boltzmann equation for the density matrix of the CMB photon ensemble are solved and it is shown that $\{\delta_{23},\delta_{32},\delta_{24},\delta_{34}\}\neq0$. These results and some other hints from recent works such as studding the interactions of CMB photons in intergalactic magnetic field \cite{Bonvin:2014xia,Giovannini} or using the effective Euler-Heisenberg Lagrangian \cite{Motie:2011az} led us to consider the above empirical form for the general scattering matrix \eqref{correction matrix}.

To proceed, let's consider a general new interaction for CMB photon. When we rotate the scattering frame to the $\hat{\mathbf{k}}=\hat{\mathbf{z}}$ frame, the correction $\delta_{ij}$ terms are represented in the scattering matrix $S(\mathbf{n},\mathbf{n}')$ as follows
\bea
S(\mathbf{n},\widetilde{\mathbf{n}})&=&S_T(\mathbf{n},\widetilde{\mathbf{n}})+S_{\textrm{New}}(\mathbf{n},\widetilde{\mathbf{n}})~,
\eea
where $S_T(\mathbf{n},\widetilde{\mathbf{n}})$ is given by \eqref{scattering matrix} and the scattering matrix due to new interactions is defined as

\beq
S_{\textrm{New}}(\mathbf{n},\widetilde{\mathbf{n}})=\frac{4\pi\tau'_{\small \textrm{New}}}{\tau'}\left(
                        \begin{array}{cccc}
                          0 & 0&0 & 0 \\
                          0 & \mathcal{S}_{22}(\mathbf{n},\widetilde{\mathbf{n}})  & \mathcal{S}_{23}(\mathbf{n},\widetilde{\mathbf{n}})   &  \mathcal{S}_{24}(\mathbf{n},\widetilde{\mathbf{n}})  \\
                          0 & \mathcal{S}_{32}(\mathbf{n},\widetilde{\mathbf{n}})   & \mathcal{S}_{33}(\mathbf{n},\widetilde{\mathbf{n}})   & \mathcal{S}_{34}(\mathbf{n},\widetilde{\mathbf{n}})  \\
                          0 &  \mathcal{S}_{42}(\mathbf{n},\widetilde{\mathbf{n}}) & \mathcal{S}_{43}(\mathbf{n},\widetilde{\mathbf{n}})  &0  \\
                        \end{array}\label{scattering matrix}
                      \right)~,
\eeq
with $\tau'_{\small \textrm{New}}$ is denoting the differential optical depth due to the new interactions of CMB photon and $\mathcal{S}_{ij}$ are the non-vanishing components in $\hat{\mathbf{k}}=\hat{\mathbf{z}}$ frame. We are now ready to calculate the imprints of the new contributions in the scattering matrix on the Boltzmann equations. Again we use the equations \eqref{Boltz0}, \eqref{scattering}, \eqref{gravityvector} and \eqref{h} and after some straightforward calculations we find the modified Boltzmann equations in the following forms
\beq
\Delta_{T}^{'S}+ik\mu\left(\Delta_{T}^{S}+\Psi\right)-\Phi'=\tau'\left(\Delta_{T0}^{S}-\Delta_{T}^{S}+ik\mu v_{B}+\frac{1}{2}P_{2}\,\Pi\right)~,
\eeq
\bea
\Delta_{Q}^{'S}+ik\mu\Delta_{Q}^{S}&=&\tau'\left(-\Delta_{Q}^{S}+\frac{1}{2}[1-P_{2}]\,\Pi\right)\nonumber \\&&+\tau'_{\textrm{ New}}\int d \widetilde{\mathbf{n}} \left[\left(\mathcal{S}_{22}+\mathcal{S}_{23}+\mathcal{S}_{32}+\mathcal{S}_{33}\right)\Delta_{Q}^{S}+
i\left(\mathcal{S}_{22}+\mathcal{S}_{32}-\mathcal{S}_{23}-\mathcal{S}_{33}\right)
\Delta_{U}^{S}\right. \nonumber \\ &&\left. \:\:\:\:\:\:\:\:\:\:\:\:\:\:\:\:\:\:\:\:\:\:\:\:\:\:\:\:\:\:\:\:\:\:\:\:+\left(\mathcal{S}_{24}+\mathcal{S}_{34}\right)\Delta_{V}^{S}\right]~,\label{neq1}
\eea
\bea
\Delta_{U}^{'S}+ik\mu\Delta_{U}^{S}=-\tau'\Delta_{U}^{S}&+&\tau'_{ \textrm{New}}\int d \widetilde{\mathbf{n}} \left[\,i\left(\mathcal{S}_{32}+\mathcal{S}_{33}-\mathcal{S}_{22}-\mathcal{S}_{23}\right)\Delta_{Q}^{S}+
\left(\mathcal{S}_{22}+\mathcal{S}_{33}-\mathcal{S}_{23}-\mathcal{S}_{32}\right)
\Delta_{U}^{S}\right. \nonumber \\ &&\left. \:\:\:\:\:\:\:\:\:\:\:\:\:\:\:\:\:\:\:\:\:\:\:\:\:\:\:\:\:\:\:\:\:\:\:\:+i\left(\mathcal{S}_{34}-\mathcal{S}_{24}\right)\Delta_{V}^{S}\right]~,\label{neq2}
\eea
\beq
\Delta_{V}^{'S}+ik\mu\Delta_{V}^{S}=-\tau'\left(\Delta_{V}^{S}+\frac{3}{2}\mu \Delta_{V1}^{S}\right)+\tau'_{ \textrm{New}}\int d \widetilde{\mathbf{n}} \left[\left(\mathcal{S}_{42}+\mathcal{S}_{43}\right)\Delta_{Q}^{S}+
i\left(\mathcal{S}_{42}-\mathcal{S}_{43}\right)
\Delta_{U}^{S}\right]~. \label{neq3}
\eeq
 As one can see the equations \eqref{neq1}, \eqref{neq2} and \eqref{neq3} are coupled to each other due to new terms in the scattering matrix. A nontrivial source for the circular polarization equation leads to a source term for the $U$ appeared on the right hand side of \eqref{neq2}. As we described above, this property is arrived in the systems with non-vanishing $V$. After a few calculations one can rewrite above differential equations \eqref{neq1}, \eqref{neq2} and \eqref{neq3} as the following integral equations form
\bea
\Delta_{V}^{S}(\eta_0,k)&=&\int_0^{\eta_0}\,d\eta\,e^{ik\mu(\eta-\eta_0)+\tau} \left(-\frac{3}{2}\tau'\mu \Delta_{V1}^{S}+\tau'_{ \textrm{New}}\int \frac{d \widetilde{\mathbf{n}}}{4\pi}\left[ \mathcal{S}_{42}\Delta^{S+}_{P}+\mathcal{S}_{43}\Delta^{S-}_{P}\right]\right)~,
\eea
\bea
   \Delta^{S+}_{P}(\eta_0,k) &=& \int_0^{\eta_0}\,d\eta\,e^{ik\mu(\eta-\eta_0)+\tau} \Big(\tau'\frac{1}{2}[1-P_{2}]\,\Pi
    +2\tau'_{ \textrm{New}}\int \frac{d \widetilde{\mathbf{n}}}{4\pi}\left[  \mathcal{S}_{22}\Delta^{S+}_{P}+\mathcal{S}_{23}\Delta^{S-}_{P}+\mathcal{S}_{24}\Delta^{S}_{V}\right]\Big)~,
\eea
\bea
\Delta^{ S-}_{P}(\eta_0,k) &=& \int_0^{\eta_0}\,d\eta\,e^{ik\mu(\eta-\eta_0)+\tau} \Big(\tau'\frac{1}{2}[1-P_{2}]\,\Pi
    +2\tau'_{ \textrm{New}}\int \frac{d \widetilde{\mathbf{n}}}{4\pi}\left[  \mathcal{S}_{32}\Delta^{S+}_{P}+\mathcal{S}_{33}\Delta^{S-}_{P}+\mathcal{S}_{34}\Delta^{S}_{V}\right]\Big),\label{Bo3}
\eea
where $\Delta^{ S\pm}_{P}=\Delta^{S}_{Q }\pm i\Delta^{S}_{U }$ and $\eta_0$ shows the present conformal time.
Using these results we now can calculate the E-mode and B-mode polarizations defined in the following manner \cite{Hu:1997hp}
\bea\label{Emode}
\Delta_{E}^{S}(\eta_0,k,\mu)&\equiv&-\frac{1}{2}[\bar{\eth}^{2}\Delta_{P}^{S+}(\eta_0,k,\mu)+\eth^{2}\Delta_{P}^{S-}(\eta_0,k,\mu)]~,
\eea
\bea
\label{Bmode}\Delta_{B}^{S}(\eta_0,k,\mu)&\equiv&\frac{i}{2}[\bar{\eth}^{2}\Delta_{P}^{S+}(\eta_0,k,\mu)-\eth^{2}\Delta_{P}^{S-}(\eta_0,k,\mu)]~,
\eea
where $\eth$ and $\bar{\eth}$ are spin raising and lowering operators respectively and we have assumed that the scalar perturbations to be axially symmetric
around ${\bf k}$ such that
\begin{eqnarray}
 \eth^{2}\, \Delta_{P}^{S\pm}=\bar{\eth}^{2}\, \Delta_{P}^{S\pm}&=&\partial_{\mu}^{2}[(1-\mu^{2})\,\, \Delta_{P}^{S\pm}(\eta_0,k,\mu)],\label{axial}
\end{eqnarray}
where here $\partial_{\mu}=\partial/\partial\mu$.
\bea\label{Emode}
\Delta_{E}^{S}(\eta_0,k,\mu)&
=&-\frac{1}{2}\int_0^{\eta_0}\,e^{\tau}d\eta\,\partial_{\mu}^{2}\Big\{(1-\mu^{2})e^{ik\mu(\eta-\eta_0)}\, \Big(\tau'\frac{1}{2}[1-P_{2}]+2\tau'_{ \textrm{New}}\int \frac{d \widetilde{\mathbf{n}}}{4\pi}\left[  (\mathcal{S}_{22}+\mathcal{S}_{32})\Delta^{S+}_{P} \right. \nonumber \\ &&\left.\:\:\:\:\:\:\:\:\:\:\:\:\:\:\:\:\:\:\:\:\:\:\:\:\:\:\:\:\:\:\:\:\:\:+(\mathcal{S}_{23}+\mathcal{S}_{33})\Delta^{S-}_{P}
+(\mathcal{S}_{24}+\mathcal{S}_{34})\Delta^{S}_{V}\right]\Big)\Big\}~,
\eea
\bea
\label{Bmode}\Delta_{B}^{S}(\eta_0,k,\mu)&
=&i\int_0^{\eta_0}\,e^{\tau}\tau'_{ \textrm{New}}d\eta\,\partial_{\mu}^{2}\Big\{(1-\mu^{2})e^{ik\mu(\eta-\eta_0)}\, \int \frac{d \widetilde{\mathbf{n}}}{4\pi}\left[  (\mathcal{S}_{22}-\mathcal{S}_{32})\Delta^{S+}_{P} \right. \nonumber \\ &&\left.\:\:\:\:\:\:\:\:\:\:\:\:\:\:\:\:\:\:\:\:\:\:\:\:\:\:\:\:\:\:\:\:\:\:\:\:\:\:
+(\mathcal{S}_{23}-\mathcal{S}_{33})\Delta^{S-}_{P}
+(\mathcal{S}_{24}-\mathcal{S}_{34})\Delta^{S}_{V}\right]\Big\}~.
\eea
The above expression for $\Delta_{B}^{S}$ is our main result. In the limit $\tau'_{ \textrm{New}}\rightarrow 0$ we recover the standard result that the scalar modes cannot generate the B-mode. For $\tau'_{ \textrm{New}}\neq 0$ the equation \eqref{Bmode} yields a minimum condition for generation of B-mode.\\
 As a result, to obtain the exact value of B-mode polarization due to Compton scattering in the case of tensor perturbations $C^{T}_B$ (as well as r-parameter), we need
to investigate all of new interactions (in additional to Compton scattering) in the case of scalar perturbations to know that 
there are any new source for B-mode polarization or not.
In the other word, the observational B-mode power spectrum $C^{ob}_B$ may generally have some new contributions from circular polarization $C^{cir}_B$ and
 also new interactions in case of scaler perturbations $C^{new,S}_{B}$ (as following) which should be determined 
\begin{equation}\label{B-mode01}
  C^{ob}_B=C^{T}_B+C^{new,S}_B+C^{cir}_B.
\end{equation}
After specifying   $C^{new,S}_B,~C^{cir}_B$, we can obtain the exact value of $C^{T}_B$ or r-parameter from experimental data
\begin{equation}\label{B-mode01}
  r\propto C^{T}_B/C^{S}_E\propto \Big[C^{ob}_B-(C^{new,S}_B+C^{cir}_B)\Big]/C^{S}_E,
\end{equation}
where $C^{S}_E$ is E-mode power spectrum  in the case of scalar perturbations.

\section{Conclusion }
Our aim is to motivate a corresponding attention to measuring the circular polarization of CMB.
Usually it is assumed that there is no physical mechanism to generate the primordial circular polarization for the CMB radiation. Hence, in the literatures the Stokes parameter $V$ is considered to be zero. However, as it has been recognized in the past years, the circular polarization mode of CMB can be non-zero due to new interactions. In this work we have generally shown that a $V\neq 0$ parameter leads to the generation of a B-mode polarization in the CMB on a background with no relic gravitational wave. In the other hand, the existing of the V-mode polarization can have effect on the value of B-mode which is detected by receive antenna. As discussed in introduction, independence of the direction of the receive antenna, the receive antenna can not cancel the measurement of circular polarization  for both cases E- and B- mode polarization in contrast to linear polarization. This means to have the exact value of B-mode (as well as r-parameter), we need the value of V-mode polarization and cancel its contribution on the B-mode.

\section*{\small Acknowledgment}

The authors would like to thank J. Khodagholizadeh for useful discussions and comments.

%%%%%%%%%%%%%%%%%%%%%%%%%%%%%%%%%%%%%%%%%%%%%%%%%%%%%%%%%%%%%%%%%%%%%%%%%%%%%%%%%%


\begin{thebibliography}{99}


\bibitem{Zaldarriaga:1996xe}
  M.~Zaldarriaga and U.~Seljak,
  ``An all sky analysis of polarization in the microwave background,''
  Phys.\ Rev.\ D {\bf 55}, 1830 (1997)
  [astro-ph/9609170].

\bibitem{Kamionkowski:1996ks}
  M.~Kamionkowski, A.~Kosowsky and A.~Stebbins,
  ``Statistics of cosmic microwave background polarization,''
  Phys.\ Rev.\ D {\bf 55}, 7368 (1997)
  [astro-ph/9611125].


\bibitem{Kosowsky:1994cy}
  A.~Kosowsky,
  ``Cosmic microwave background polarization,''
  Annals Phys.\  {\bf 246}, 49 (1996)
  [astro-ph/9501045].


\bibitem{Hu:1997hp}
  W.~Hu and M.~J.~White,
  ``CMB anisotropies: Total angular momentum method,''
  Phys.\ Rev.\ D {\bf 56}, 596 (1997)
  [astro-ph/9702170].
%%%%%%%%%%%%%%%%%%%%%%%%%%%%%%%%%%%%%%%%%%%%%%%%%
\bibitem{Guth:1980zm}
  A.~H.~Guth,
  ``The Inflationary Universe: A Possible Solution to the Horizon and Flatness Problems,''
  Phys.\ Rev.\ D {\bf 23}, 347 (1981).

\bibitem{Linde:1981mu}
  A.~D.~Linde,
  ``A New Inflationary Universe Scenario: A Possible Solution of the Horizon, Flatness, Homogeneity, Isotropy and Primordial Monopole Problems,''
  Phys.\ Lett.\ B {\bf 108}, 389 (1982).

 \bibitem{Linde:1983gd}
  A.~D.~Linde,
  ``Chaotic Inflation,''
  Phys.\ Lett.\ B {\bf 129}, 177 (1983).

  \bibitem{Bardeen:1983qw}
  J.~M.~Bardeen, P.~J.~Steinhardt and M.~S.~Turner,
  ``Spontaneous Creation of Almost Scale - Free Density Perturbations in an Inflationary Universe,''
  Phys.\ Rev.\ D {\bf 28}, 679 (1983).


 \bibitem{Kosowsky:1995aa}
  A.~Kosowsky and M.~S.~Turner,
  ``CBR anisotropy and the running of the scalar spectral index,''
  Phys.\ Rev.\ D {\bf 52}, 1739 (1995)
  [astro-ph/9504071].



%%%%%%%%%%%%%%%%%%%%%%%%%%%%%%%%%%%%%%%%%%%%%%%%%
  \bibitem{Planck}
  P.~A.~R.~Ade {\it et al.}  [Planck Collaboration],
  ``Planck 2015. XX. Constraints on inflation,''
  arXiv:1502.02114 [astro-ph.CO].
  %%CITATION = ARXIV:1502.02114;%%
  %34 citations counted in INSPIRE as of 14 Mar 2015

%%%%%%%%%%%%%%%%%%%%%%%%%%%%%%%%%%%%%%%%%%%%%%%%%%%%%%%%%%

\bibitem{Bonvin:2014xia}
 C.~Bonvin, R.~Durrer and R.~Maartens,
   ``Can primordial magnetic fields be the origin of the BICEP2 data?,''
 [arXiv:1403.6768 [astro-ph.CO]].

\bibitem{Khodagholizadeh:2014nfa}
  J.~Khodagholizadeh, R.~Mohammadi and S.~S.~Xue,
  ``Photon-neutrino scattering and the B-mode spectrum of CMB photons,''
  Phys.\ Rev.\ D {\bf 90}, no. 9, 091301 (2014)
  [arXiv:1406.6213 [astro-ph.CO]].

\bibitem{Moss:2014cra}
  A.~Moss and L.~Pogosian,
  ``Did BICEP2 see vector modes? First B-mode constraints on cosmic defects,''
  Phys.\ Rev.\ Lett.\  {\bf 112}, 171302 (2014)
  [arXiv:1403.6105 [astro-ph.CO]].

  \bibitem{Lizarraga:2014eaa}
  J.~Lizarraga, J.~Urrestilla, D.~Daverio, M.~Hindmarsh, M.~Kunz and A.~R.~Liddle,
  ``Can topological defects mimic the BICEP2 B-mode signal?,''
  Phys.\ Rev.\ Lett.\  {\bf 112}, 171301 (2014)
  [arXiv:1403.4924 [astro-ph.CO]].


  %%%%%%%%%%
\bibitem{Mohammadi:2013}
  R.~Mohammadi,
  ``Evidence for cosmic neutrino background form CMB circular polarization,''
 Eur.\ Phys.\ J.\ C  74:3102(2014), arXiv:1312.2199 [astro-ph.CO].

  \bibitem{Alexander:2008fp}
  S.~Alexander, J.~Ochoa and A.~Kosowsky,
  ``Generation of Circular Polarization of the Cosmic Microwave Background,''
  Phys.\ Rev.\ D {\bf 79}, 063524 (2009)
  [arXiv:0810.2355 [astro-ph]].

  \bibitem{Bavarsad:2009hm}
  E.~Bavarsad, M.~Haghighat, Z.~Rezaei, R.~Mohammadi, I.~Motie and M.~Zarei,
  ``Generation of circular polarization of the CMB,''
  Phys.\ Rev.\ D {\bf 81}, 084035 (2010),
  [arXiv:0912.2993 [hep-th]].


\bibitem{Giovannini}
M~. Giovannini, ``Last scattering, relic gravitons and the circular polarization of the CMB,''
 Phys.\ Rev.\ D {\bf 81},123003 (2010)
[arXiv:1001.4172 [astro-ph.CO]].


\bibitem{Motie:2011az}
  I.~Motie and S.~S.~Xue,
  ``Euler-Heisenberg Lagrangian and Photon Circular Polarization,''
  Europhys.\ Lett.\  {\bf 100}, 17006 (2012)
  [arXiv:1104.3555 [hep-ph]].




 \bibitem{De:2014qza}
  S.~De and H.~Tashiro,
  ``Circular Polarization of the CMB: A probe of the First stars,''
  arXiv:1401.1371 [astro-ph.CO].


\bibitem{Sawyer:2014maa}
  R.~F.~Sawyer,
  ``Photon-photon interactions can be a source of CMB circular polarization,''
  arXiv:1408.5434 [astro-ph.CO].

  \bibitem{Mainini:2013mja}
  R.~Mainini, D.~Minelli, M.~Gervasi, G.~Boella, G.~Sironi, A.~Baú, S.~Banfi and A.~Passerini {\it et al.},
  ``An improved upper limit to the CMB circular polarization at large angular scales,''
  JCAP {\bf 1308}, 033 (2013)
  [arXiv:1307.6090 [astro-ph.CO]].

  %%%%%%%%%%%%%%%%%%%
  \bibitem{BICEP2}
 P.~A.~R.~Ade {\it et al.}  [BICEP2 Collaboration],
 ``BICEP2 I: Detection Of B-mode Polarization at Degree Angular Scales,''[arXiv:1403.3985 [astro-ph.CO]];
 %%%%%%%%%%%%%%%%%%%%%%%%%%%%%%%%%%%%
\bibitem{Chandrasekhar} S. Chandrasekhar, ``Radiative Transfer'', Dover,
New York, 1960.




%%%%%%%%%%%%%%%%%%%%%%%%%%%%%%%%%%%%%%%%%%%%%%%%%%%%%%%%%%%









\end{thebibliography}
\end{document}